# Laser cooling of solids

Galina Nemova

*Ecole Polytechnique de Montréal*
*Canada*

## 1. Introduction

At the present the term "laser cooling", which does not imply the cooling of lasers, is most often used to describe cooling and trapping of atoms and ions to extremely low temperatures right to the nanokelvin regime. There are differant methods of laser cooling of atoms amd ions, for example, Doppler cooling (Hänsch & Schawlow, 1975), where light forces are exerted by absorption and subsequent spontaneius emission of photons. The rate of these processes depends on the velocity of an atom or ion due to the Doppler shift. This method is limited in terms of the reachable temperature by *Doppler limit*. Sisyphus cooling (Dalibard & Cohen-Tannoudji, 1985), which involves a polarization gradient and therefore sometimes is called polarization gradient cooling, allows to get temperatutes substantially below the Doppler limit. It goes down to the much lower *recoil limit* associated with the recoil momentum related to the absorption or emission of a single photon. But the recoil limit can be overcamed by the method of velocity-selective coherent population trapping (Aspect et al., 1988), which allows *sub-recoil temperatures* in the nanokelvin regime. This area of sience has progressed immensely in the last two decades facilitating the observation of Bose-Einstin condensates and many related phenomena and resulting in physics Nobel prizrs 1997 (Chu, Cohen-Tannoudji & Philips, 1997) and 2001 (Cornell, Ketterle & Weiman, 2001).

It is worth to emphasise, what may be many researchers don't realize, that nearly half a cenury before Doppler cooling of atoms was observed, cooling of solids by anty-Stoks fluorescence was proposed by Peter Pringsheim (Pringsheim, 1929). It has been shown that some materials emitted light at shorter wavelengths than that with which the material was illuminated due to thermal (phonon) interactions with the excited atoms. This process was named anti-Stokes fluorescence opposed to the Stokes fluorescence in which the emitted wavelength is greater than the absorbed one. Since anti-Stokes fluorescence involves the emission of higher energy photons than those which were absorbed, the net anti-Stokes fluorescence can cause removal of energy from the material illuminated and, as a consequence, its refrigeration. Of course, it is essential that the qauntum efficiency is high, and that nearly all anti-Stokes fluorescence light can leave the crystal without being re-absorbed. Contrary to the gas phase, where particals (moleculs, atoms, ions, electrons, etc.) are in random motion, in the solid phase, atoms do not have relative translational motion. In solids thermal energy is largely contained in the vibrational modes of the lattice (phonons). Laser cooling of solids (also called optical, laser refrigeration or anti-Stokes fluorescent cooling) is similar to laser cooling of atoms and ions: photons in the red tail of the absotption spectrum are absorbed from a monochromatic source followed by anti-Stokes fluoresce of blue-shifted photons with higther energy. These anti-Stokes potons remove the phonons from the solid and cool it.

This idea optical cooling of solids stimulated a discussion between Vavilov and Pringsheim concerning thermodynamic aspect of this process. Vavilov believed (Vavilov, 1945 and 1946) that optical cooling of solids by anti-Stokes fluorescence contradicts the second law of thermodynamics. Vavilov argued that the cycle, which includes excitation and fluorescence, is a reversible one and the energy yield of greater than unity would be equivalent to the complete transformation of heat to work. Prigsheim contend (Pringsheim, 1946) that in anti-Stokes cooling, a monochromatic and unidirectional input beam transforms into isotropic broadband fluorescence, and therefore this process must be irreversible. In 1946, Landau (Landau, 1946) concluded this controversy by providing a thermodynamic theory of the process by considering the entropy of the incident and scattered light. Using Bose statistics on the photon gas and integrating over the solid angle and bandwidth of the radiation, he calculated an effective temperature for the radiation concluding that it is possible for the energy yield to exceed unity. In 1950, Kastler (Kastler, 1957) and in 1961, Yatsiv (Yatsiv, 1961) suggested to use rare-earth (RE) doped solids for laser cooling. The main advantage of RE-ions is the optically active $4f$ electrons shielded by the filled $5s$ and $5p$ outer shells, which limit interaction with the lattice surrounding the RE-ion and suppress non-radiative decay. Hosts with low phonon energy, for example, fluoride glasses and crystals, can diminish non-raditive decay and increase quantum efficiency. In 1968, Kushida and Geusic (Kushida & Geusic, 1968) in attempt to cool a $Nd^{3+}$:YAG crystal observed a reduction of heating. Net radiation cooling by anti-Stokes fluorescence in solid materials was observed experimentally for the first time only in 1995 by Epstein's research team at Las Alomos National Laboratory (LANL) in ytterbium-doped fluorozirconate $ZrF_4$-$BaF_2$-$LaF_3$-$AlF_3$-$NaF$-$PbF_2$ (ZBLANP) glass (Epstein et al., 1995). A $Yb^{3+}$-doped sample of ZBLANP in the shape of rectangular parallelepiped of volume 43 $mm^3$, was cooled to 0.3 K below room temperature.

Laser cooling of solids at the present time can be largely divided into three areas: laser cooling with ion doped glasses or crystals, laser cooling in semiconductors (bulk or confined, like quantum-well structures), and radiation-balanced lasers, where the pump wavelength is adjusted so that the anti-Stokes fluorescence cooling compensates for the laser heating. The first two directions are devoted to development of an all-solid-state cryalooler. Being entirely solid state, optical coolers have the advantages of being compact, the absence of mechanical vibrations, moving parts or fluids. They have many potential benefits over thermoelectric and mechanical coolers. For example, optical refrigerators share the benefit of low mechanical vibrations with thermoelectric coolers (TEC) based on Peltier effect, but do not suffer from the need to remove heat by physical contact with a heat-sink. TECs are able to reach ~180 K only and thus Peltier coolers are more effective than optical coolers for high temperature operations (> 190 K), while anti-Stokes coolers can cool at temperatures as low as ~80 K (for RE-doped glasses or crystals) or ~55 K (for direct-bandgap semiconductors). Mechanical coolers such as the Stirling cycle cooler can reach temperatures of order 10 K but they are large and cause vibrations that may not be suitable in many applications. Optical coolers can be based on long-lived diode laser systems, remote from the cooler, and cause low electromagnetic interference in the cooling area. All three areas are equally important and promising.

## 2. Laser cooling in ion-doped glasses and crystals

For decades the advantages of laser cooling of solids were connected with rare-earth (RE) doped glasses and crystals. As mentioned above, the main advantage of RE-ions is the optically active 4*f* electrons shielded by the filled 5*s* and 5*p* outer shells, which limit interaction with the lattice surrounding the RE-ion and suppress non-radiative decay. Hosts with low phonon energy, for example, fluoride glasses and crystals, can diminish non-raditive decay and increase quantum efficiency. Laser-induced cooling has been observed in a wide variety of glasses and crystals doped with ytterbium ($Yb^{3+}$) such as ZBLANP (Epstein et al., 1995, Mungan et al., 1997, Luo et al., 1998, Gosnell, 1999 & Thiede et al., 2005), ZBLAN (Rayner et al., 2001 & Heeg et al., 2004), CNBZn and BIG (Fernandez et al., 2000), YAG and $Y_2SiO_5$ (Epstein et al., 2001), $BaY_2F_8$ (Bigotta et al., 2006), $KPb_2Cl_5$ (Mendioroz et al., 2002), $KGd(WO_4)_2$ and $KY(WO_4)_2$ (Bowman et al., 2000), YLF (Seletskiy et al., 2008). Laser-induced cooling has been also observed in thulium ($Tm^{3+}$) doped ZBLANP (Hoyt et al., 2000 & 2003) and $BaY_2F_8$ (Patterson et al., 2008), and in erbium ($Er^{3+}$) doped CNBZn and $KPb_2Cl_5$ (Fernandez et al., 2006 & Garcia-Adeva et al., 2009).

In 1963, Tsujikawa and Murao proposed that running a ruby laser, which is based on transition metal doped host ($Cr^{3+}$-doped ruby), in reverse might result in cooling (Tsujikawa & Murao, 1963). Later spectroscopic analysis of $Cr^{3+}$-doped ruby shows Stokes sidebands, which would completely suppress any cooling effects (Nelson & Sturge, 1965). In 2008, Ruan and Kaviany developed an *ab intio* approach to determine the photon-electron and electron-vibration coupling rates, for ion-doped materials related to laser cooling of solids (Ruan & Kaviany, 2008). They used $Ti[(H_2O)_6]^{3+}$ complex rather than a periodic solids, because the excited states of a finite system is much simpler to analyze while the physics is preserved. It has been shown that the phonon-assisted absorption process is the slowest process in cooling cycle and regarded as the bottleneck of laser cooling of solids. To enhance the electron-phonon coupling the modification of vibrational modes before and after transition should be significant. Contrary to rare-earth ions, where valence electrons are shielded by outer electrons, transition metal ions have its valence electrons directly exposed to the crystal field. As a result, transition-metal doped hosts have a much larger electron-phonon coupling. A negative side effect of the electron-phonon coupling is the nonradiative decay. However, at very low cryogenic temperatures the nonradiative decay rate is small. We suppouse the transition metal doped host can be used for cooling at very low temperatures. Laser cooling in solids doped with transition metal has not been observed yet.

**2.1 The 4-level model for optical refrigeration**
Consider basic concepts of laser cooling of solids using as an example of $Yb^{3+}$:ZBLANP sample. Energy levels in *cm*$^{-1}$ and major transitions of $Yb^{3+}$ in ZBLANP are illustrated in Fig. 1 (a). We will approximate the systems of levels illustrated in Fig. 1 (a) by the 4-level system illustrated in Fig. 1 (b). In this 4-level system the ground state manifold ($^2F_{7/2}$) is presented by two energy levels with an energy separation $\delta E_g = E_1 - E_0$, corresponding to the bottom ($E_0$) and to the top ($E_1$) of this manifold. The exited manifold ($^2F_{5/2}$) is presented by two energy levels with an energy separation $\delta E_{ex} = E_3 - E_2$, corresponding to the bottom ($E_2$) and

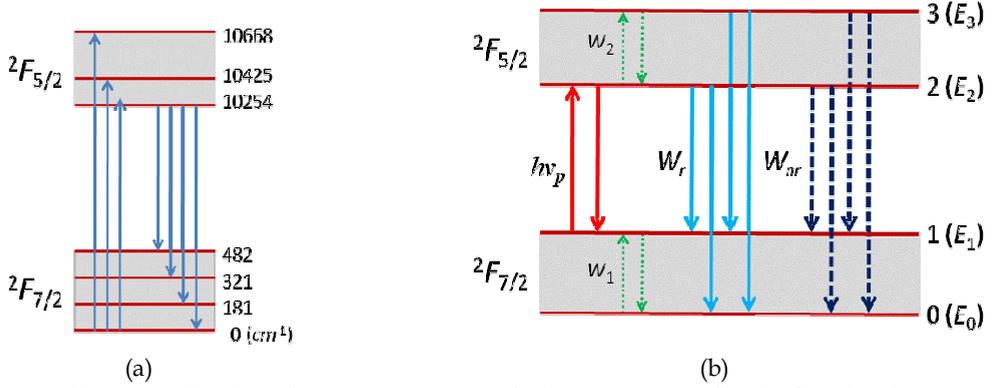

Fig. 1. (a) Energy levels and major transitions of Yb$^{3+}$ in ZBLAN. (b) The four-level energy model for optical refrigeration consisting of two pairs of levels in the ground (0 and 1) and exited (2 and 3) manifolds.

to the top ($E_3$) of this manifold too. The pump signal at the frequency $v_p$ is in resonance with the 1 -> 2 transition. We assume that all spontaneous transitions from 3 and 2 levels down to 1 and 0 levels have the same rate $W_r$. We also assume that the rate of nonradiative transitions between these levels illustrated by the dashed arrows in Fig. 1 (b) is equal to $W_{nr}$. Thermolization, in the ground and exited manifolds, is illustrated by dotted arrows. It is based on electron-phonon interation, which has rates $w_g$ and $w_{ex}$ for lower and exited manifolds, respectively. The evolution of the density populations of the levels can be described by the rate equations:

$$\frac{dN_1}{dt} = -\sigma_{12}(N_1 - N_2)\frac{I}{h\nu_p} - N_1 w_g + (N_2 + N_3)(W_r + W_{nr}) + N_0 w_g,$$

$$\frac{dN_2}{dt} = \sigma_{12}(N_1 - N_2)\frac{I}{h\nu_p} - 2N_2(W_r + W_{nr}) + \left(N_3 - N_2 \exp\left(-\frac{\delta E_{ex}}{k_B T}\right)\right)w_{ex},$$

$$\frac{dN_3}{dt} = -\left(N_3 - N_2 \exp\left(-\frac{\delta E_{ex}}{k_B T}\right)\right)w_{ex} - 2N_3(W_r + W_{nr}),$$

$$N = N_0 + N_1 + N_2 + N_3,$$

(1)

where $N_i$ is the population of the levels $i$th, $i$ = 1, 2, 3, and 4. $\sigma_{12}$ is the absorption cross section corresponding to 1->2 transition. $I$ is the intensity of the pump signal. $k_B$ is the Boltzmann's constant. We assume that degeneracy for all levels is equal to unity. The difference between absorbed and radiated power densities is equal to

$$P_{net} = [\sigma_{12}N_1 + \alpha_b]I - \{[(N_2 + N_3)(2h\nu_p + \delta E_g) + 2N_3\delta E_{ex}]W_r + \sigma_{12}N_2 I\}, \quad (2)$$

where $\alpha_b$ is the coefficient of the parasitic absorption. This is the net power density deposited in the system. We will ignore saturation and consider the steady state condition equating all derivatives in equation (1) to zero. Taking into account the internal quantum efficiency $\eta_q = W_r/(W_r+W_{nr})$, and introducing the mean fluorescence energy as

$$h\nu_f = h\nu + \frac{\delta E_g}{2} + \frac{\delta E_{ex}}{1 + \left(\frac{2(W_r + W_{nr})}{w_{ex}} + 1\right)\exp\left(\frac{\delta E_g}{k_B T}\right)}, \qquad (3)$$

one can re-write the equation for the net power density deposited in the system in the form:

$$P_{net} = (\alpha_{res} + \alpha_b)I - \alpha_{res}\eta_q \frac{\nu_f}{\nu_p} I, \qquad (4)$$

where $\alpha_{res}$ is the ground state absorption.

$$\alpha_{res} = \frac{\sigma_{12} N}{(1 + \exp(\delta E_g/k_B T))}. \qquad (5)$$

It exhibits diminishing pump absorption caused by thermal depletion of the top ground state with decries of the temperature, $k_B T < \delta E_g$. In order to achieve cooling at low temperatures with reasonable efficiency the width of the ground manifold, $\delta E_g$, has to be narrow. As one can see in Eq. (3) the mean fluorescence photon energy becomes red shifted with decrease in the temperature. These decreases further the efficiency of the cooling. If $w_2 < (W_r+W_{nr})$, where the $W_r+W_{nr}$ is the total upper state decay rate, the decay of the excited state occurs before thermalization with the lattice and no cooling will result. The total absorption power density is $P_{abs} = (\alpha_{res}+\alpha_b)I$. Using (4) one represent the cooling efficiency, $\eta_{cool} = \Box P_{net}/P_{abs}$, as

$$\eta_{cool} = \eta_q \eta_{abs} \frac{\nu_f}{\nu_p} - 1, \qquad (6)$$

where $\eta_{abs} = \alpha_{res}/(\alpha_{res}+\alpha_b)$ is an absorption efficiency defining the fraction of pump photons involved in cooling process. The equation (6) does not take in to account the fluorescence trapping and re-absorption. If one wants to taken into account these effects $\eta_q$ in the equation (6) has to be replaced by an external quantum efficiency $\eta_{ext} = \eta_e W_r/(\eta_e W_r+W_{nr})$, where $\eta_e$ is a fluorescence escape efficiency. The minimum temperature, which can be archived can be calculated from the equation $\eta_{cool} = 0$. Increasing the purity of the host, choosing a system with narrow ground manifold, increasing the quantum efficiency, and preventing the trapping and re-absorption one can decrease the value of this minimum temperature.

**2.2 The 2-level model for optical refrigeration**

Basic concepts of laser cooling of solids can be described using a two level model for absorption and stimulated-emission processes between the Yb$^{3+}$ ground-state manifold, $^2F_{7/2}$, and the exited-state, $^2F_{5/2}$, manifold too. This model is based on two equations. The first one describes the total steady-state population density in the excited-state manifold, $N_{ex}$, have to satisfy the equation

$$\frac{dN_{ex}}{dt} = \frac{I}{h\nu_p}(\sigma_{abs}N_g - \sigma_{se}N_{ex}) - \eta_e N_{ex}W_r - N_{ex}W_{nr} = 0, \qquad (7)$$

where $N_g + N_{ex} = N$ is the total number density of ions, $I$ is the pump intensity at the pump wavelength $\lambda_p = c/\nu_p$. $\sigma_{abs}$ and $\sigma_{se}$ are the absorption and the stimulated emission cross sections, respectively, at the pump wavelength. $W_r = 1/\tau_r$ and $W_{nr} = 1/\tau_{nr}$ are the radiative and nonradiative rates, respectively, which are the inverses of the lifetimes.

The second equations describing a two level system is the rate of thermal energy, $u$, accumulation in the volume element of the sample:

$$\frac{du}{dt} = \sigma_{abs}N_g I - \sigma_{se}N_{ex}L + \alpha_b I - \eta_e N_{ex}h\nu_f W_R, \qquad (8)$$

The negative rate of the thermal energy accumulation defines the cooling power density, $P_{cool} = -du/dt$. The relative cooling efficiency can be defined as $\eta_{cool} = P_{cool}/P_{abs}$. $\nu_f = c/\lambda_f$ is a mean fluorescence frequency connected with the mean fluorescence wavelength, $\lambda_f$, which can be calculated from the measured fluorescence spectrum including the red shifting due to reabsorption with the equation $\lambda_f = (\int \lambda F_\lambda d\lambda)(\int F_\lambda d\lambda)^{-1}$. Using equations (7) and (8) one can present $\eta_{cool}$ in the form

$$\eta_{cool} = \frac{\sigma_{abs}(1 - \lambda/\lambda_{f^*})I_s/I}{1 + \sigma_{se}/\sigma_{abs} + I_s/I}, \qquad (9)$$

where $I_s = hc\eta_e W_r/(\lambda \sigma_{abs}\eta_{ext})$ is a wavelength-dependent saturation intensity and $\lambda_{f^*} = [\eta_{ext}(1/\lambda_f - \alpha_b/(hc\eta_e W_r))]^{-1}$ is the effective mean fluorescence wavelength. If the wavelength of the pump signal is equal to the effective mean fluorescence wavelength neither heating nor cooling will be observed. Pumping at the wavelength longer than the effective mean fluorescence wavelength will causes net anti-Stokes fluorescence and possible optical refrigeration. Indeed in this case the escaping light carries more energy than the absorbed pump light. At very long wavelengths absorption by impurities or imperfections dominates, and the sample heats.

It is important to emphasise that the appropriate choice of laser cooling medium is a very important factor if one wants to get high efficient laser cooling of solid. The rate of non-radiative multi-phonon decay resulting to heat generation in the host decreases exponentially with the separation between the ground and exited manifolds

$$W_{nr} = W_0 \exp(-\alpha \Delta E). \qquad (10)$$

This is well-known energy-gap law. $W_0$ is a phenomenological parameter that depends strongly on host material. The parameter $\alpha$ is inversely proportional to the characteristic phonon energy in a given material. It is strongly host-dependent. The $\Delta E$ is the energy gap of the particular electronic state transition.

## 2.3 Optical refrigeration with upconversion

As is seen from previous paragraphs in an anti-Stokes cycle the energy of the fluorescent photons is slightly larger than the energy of the incident one and a small amount of the thermal vibrational energy is removed from the sample. Apart from this conventional cooling cycle the upconversion process in the host with low phonon energy can cause cooling of a system. Consider basic concepts of laser cooling of solids by upconversion using as an example of $Er^{3+}$:$KPb_2Cl_3$ sample. The upconversion cooling process is illustrated in Fig. 2, where energy levels of an $Er^{3+}$-doped $KPb_2Cl_3$ crystal are presented. The $^4I_{9/2}$ level is pumped with CW laser below the barycentre of the level. This level has long lifetime (2.4 ms) and an electronic population reservoir is created in this level. Part of this population decays spontaneously to the level $^4I_{15/2}$ by means of a direct transition or through a sequential one involving level $^4I_{13/2}$. Another part of this population is involved in upconversion process, which has two channels: the excited-state absorption (ESA), and energy transfer upconversion (ETU). In the case of the ESA an electron in the exited state is promoted in to a higher excited level by another pumping photon from where it later decays spontaneously. In the case of the ETU two electrons in the exited states of two different ions interact with each other and, as a result, one of them decays to the ground state whereas the other one is promoted to a higher excited state, from where it later decays spontaneously to the ground state. The population dynamic of the electronic levels is described using rate equations.

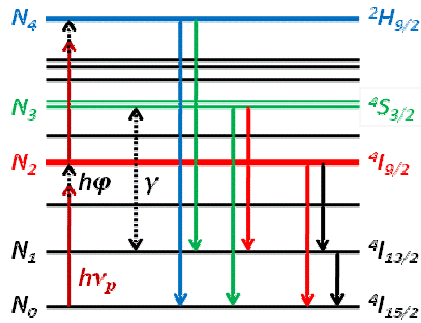

Fig. 2. Energy levels and major transitions of $Er^{3+}$ in $KPb_2Cl_5$ host.

$$\frac{dN_1}{dt} = -\frac{N_1}{\tau_1^{rad}} + \frac{N_2}{\tau_2^{rad}} + \gamma N_2^2 + \frac{N_3}{\tau_3^{rad}} + \frac{N_4}{\tau_4^{rad}},$$

$$\frac{dN_2}{dt} = \frac{I}{h\nu_p}\alpha_r - 2\frac{N_2}{\tau_2^{rad}} - 2\gamma N_2^2 - \sigma_{ESA}\frac{I}{h\nu_p}N_2,$$

$$\frac{dN_3}{dt} = -2\frac{N_3}{\tau_3^{rad}} + \gamma N_2^2, \qquad (11)$$

$$\frac{dN_4}{dt} = \sigma_{ESA}\frac{I}{h\nu_p}N_2 - 2\frac{N_4}{\tau_4^{rad}},$$

$$N = N_0 + N_1 + N_2 + N_3 + N_4,$$

where $N_i$ ($i=0,...,4$) and $\tau_i^{rad}$ are the populations and the intrinsic lifetimes of the $i$th level, respectively. $N$ is the concentration of the active ions in the host. $I$ is the intensity of the pump signal, $\gamma$ is the strength of the ETU process, $\sigma_{ESA}$ is the ESA cross section, and $\nu_p$ is the frequency of the pump signal. $\alpha_r$ is the resonant absorption coefficient. We consider a very low phonon host and for simplicity the nonradiative decays are not taken into account. For simplicity we assume that the traction of photons escaping the sample, which depends from the geometry of sample, its refractive index, etc., is almost unity ($\eta_e^{(i)} \approx 1$). A steady-state condition, $dN_i/dt = 0$, ($i = 1,..,4$), is considered. The net power density deposited in the sample can be calculated with the equation

$$P_{net} = \alpha I - \frac{h\overline{\nu_2}}{\tau_2^{rad}}N_2 - \frac{h\overline{\nu_3}}{\tau_3^{rad}}N_3 - \frac{h\overline{\nu_4}}{\tau_4^{rad}}N_4. \qquad (12)$$

where $\alpha = \alpha_r + \alpha_b$, and $\alpha_b$ is a background absorption coefficient. $\overline{\nu_i}$ is the mean fluorescence frequency of the $i$th emitting band. In this expression, the first three terms on the right side describe the conventional anti-Stokes cooling mechanism. The fourth and fifth terms describe cooling with upconversion through the ESA and the ETU channels, respectively. Dividing the Eq. (12) by the total absorbed power density $P_{abs} = \alpha I$ one can obtain the cooling efficiency $\eta_c = -P_{net}/P_{abs}$. It can be expressed as

$$\eta_{cool} = \frac{1}{\alpha}\left[\frac{\alpha_r\frac{\overline{\nu_2}}{\nu_p}}{2+\tau_2^{rad}\sigma_{ESA}\frac{I}{h\nu_p}} + \frac{\gamma}{2}h\overline{\nu_3}\frac{\alpha_r^2\frac{I}{(h\nu_p)^2}}{\left(\frac{2}{\tau_2^{rad}}+\sigma_{ESA}\frac{I}{h\nu_p}\right)^2} + \frac{\sigma_{ESA}}{2}\frac{\alpha_r h\overline{\nu_4}\frac{I}{(h\nu_p)^2}}{\left(\frac{2}{\tau_2^{rad}}+\sigma_{ESA}\frac{I}{h\nu_p}\right)}\right] - 1. \qquad (13)$$

This expression shows that both upconversion channels, the ESA and the ETU, are involved in upconversion cooling. Illuminating upconversion and setting $\gamma = \sigma_{ESA} = 0$ in Eq. (13), one can obtain the conventional model for laser cooling with anti-Stokes fluorescence. Setting $\tau_2{}^{rad}$->∞, one can neglect the standard cooling cycle as compared to upconversion cooling. It is important to emphasise that contrary to conventional cooling with anti-Stokes fluorescence, which was considered in privies paragraphs, the upconversion cooling has an *intensity threshold*, that is, the minimum input intensity of the pump beam has to be provided to get cooling. This intensity threshold can be easily calculated with Eq. (13). It is equal to

$$I_p^{th} = h\overline{\nu_3} \frac{\gamma \overline{\alpha_r}^2}{2\alpha \sigma_{ESA}^2} \left( \frac{\overline{\alpha_r}}{2\alpha} \frac{\overline{\nu_4}}{\nu_p} - 1 \right)^{-1}. \qquad (14)$$

The results of the paragraph show that upconversion cooling can be used in the development of a solid-state cryocooler.

**2.4 Laser cooling of rare-earth doped solids**
As was already mentioned laser cooling of solid has been observed in a number of glasses and crystals doped with $Yb^{3+}$, $Tm^{3+}$, and $Er^{3+}$ ions. The appropriate choice of the host is the most significant factor for laser cooling of solids. The materials with low maximum phonon energies, such as ZBPLANP or other fluorides, where non-radiative decays are relatively slow, are the best choice for the host. The sulphide- and chloride-based hosts have lower the maximum phonon energy than that of fluoride compounds. It causes a significant reduction of the multiphonon relaxation rates. However, these materials are difficult to synthesize and have poor mechanical properties. In hosts, such as oxide crystals or glasses, rapid non-radiative decay rates can prevent laser cooling. The purity of the host material is very important factor for laser cooling process, since quenching of excited ions by impurities, such as iron and copper, serves as an important source of heating in the material. A recent study shows that the almost ideal cooling efficiency in $Yb^{3+}$-doped fluorozirconate glass ZBLAN can be achieved when concentration of impurities such as $Cu^{2+}$ is less than 0.01 ppm and that for $Fe^{2+}$ is less than 0.1 ppm (Hehlen et al., 2007). The shape of the sample is the second important factor, which has to be taken into account if one wants to maximize the temperature drop of a sample. The trapping of anti-Stokes fluorescence and its re-absorption serves as an additional source of parasitic heating and has to be minimised. The best candidate for the cooled sample to meet the criterion would be a small sphere. But the spherical shape of the cooling sample would significantly limit the amount of absorbed pump power. A geometry in which the pump beam can travel for a long distance through the sample, but fluorescence still exists within a short distance is preferable for laser cooling of solids. The calculation of the external quantum efficiency caused by fluorescence re-absorption and re-emission for the samples with the cylinder and cuboid geometries using Monte-Carlo method have shown that cylinder (fiber) geometry is the best choice for the cooled sample (Wang et al., 2007). The radius of the incident pump beam has to be much less than the radius of the fiber sample, and propagates along its axis. The radius of the fiber has to be optimized for the value of the pump power (Nemova & Rashyap, 2008 & 2009).

      There are also another ways to improve the laser cooling process. For example, multipassing of the pump beam can enhance the absorbed pump radiation. It is important to emphasise that in the case of multipassing the reflected light should be as collinear as

possible in order to minimize re-absorption. Multipassing permitted decrease the cooling temperature as one can see in Refs. (Gosnell, 1999 & Edwards et al., 1999). Multipassing can be realized with an extra-cavity (EC) and intra-cavity (IC) schemes. In the case of EC scheme the cooling sample is separated from the pump laser and bricked up between two mirrors, which trap the pump beam until it is absorbed. In the case of IC scheme the cooling sample is placed inside the laser resonator with its inherent colinear multipassing. These schemes are different. For example, in the EC scheme an increase in the optical density causes an increase in the absorbed power. In the case of the IC scheme the cooling sample absorbing the power for optical cooling induces an additional loss inside the resonator and lowers the intra-cavity power. Changing the optical density in the IC scheme one has to change the condition of laser oscillation. Detailed comparison between these two schemes has been done by Heeg and Rumbles (Heeg & Rumbles, 2002). It is based on the attenuation of light described with Beer's law. It is assumed that the geometers of the samples are similar and the reabsorption of anti-Stokes fluorescence affects the cooling efficiency in a similar fashion in both configurations. It has also been assumed that the thermal lenses are comparable magnitude in both schemes, although the lensing is proportional to the profile of the beam intensity and has to be different in these configurations. The IC scheme is preferred for low optical densities (<0.01), and the EC scheme is preferred for higher optical densities (>0.1).

Ruan and Kaviany have theoretically demonstrated that the cooling efficiency can be dramatically enhanced then host material containes rare-ion-doped nanocrystalline powder (Ruan & Kaviany, 2006). Light localization effects increase the photon density, leading to an enhanced absorptivity and the phonon spectrum is modified due to finite size of the nanopowder. Later it has been shown that using a photonic crystal doped with RE ions offers many advantages with regards to getting a larger cooling efficiency at room temperature when compared to standard bulk materials or nano-powders (Garcia-Adeva et al, 2009). This structire permits to control the spontaneous emission rate of the ions embedded in the structure and the absorption rate in the Stokes side of the absorption spectrum by adequately tuning the density of photonic states, and as a consiquence obtaining a large improvement in the cooling efficiency.

Finely we want to mention the main stages of experimental development of laser cooling of solids in RE-doped glasses and crystals. For the first time laser cooling of solids was observed in a proof-of-principle experiment with $Yb^{3+}$:ZBLANP (Epstein, 1995). $Yb^{3+}$:ZBLANP is a two-level system free from the ESA. Laser cooling of $Tm^{3+}$-doped ZBLANP was the first demonstration of optical refrigeration in the presence of the ESA, although ESA had not been used as a mechanism of the cooling process. The ground and the first exited manifolds were involved in cooling cycle in the case of thulium doped system (Hoyt et al, 2000). Laser cooled $Tm^{3+}$:ZBLANP demonstrated nearly a factor of two encasement in the cooling efficiency compared to $Yb^{3+}$:ZBLANP. In 2006, $Er^{3+}$:$Kpb_2Cl_5$ crystal and $Er^{3+}$:CNBZn glass were optically cooled with infrared-to-visible upconversion. Erbium doped samples were pumped at $\lambda \sim 870$ nm. The reduced multiphonon transition rates of the $Kpb_2Cl_5$ and CNBZn hosts make possible for the pump level $^4I_{9/2}$ with long lifetime (2.4 ms) to act as an intermediate photon reservoir from which additional upconversion process take place through ESA and ETU channels. Upconversion opens new possibilities for optical cooling of solids.

## 3. Optical cooling in semiconductors

The recent advances in development and fabrication of semiconductor lasers have stimulated growth of an interest to semiconductors as candidates for optical cooling. The essential difference between semiconductors and rare-earth doped materials is in their cooling cycles. In the case of RE-doped glasses or cryctals, the cooling transition occurs in localized donor ions within the host. In the case of semiconductors, the cooling cycle involves transition between extended valence and conduction bands of a direct band gap semiconductor (Fig. 3). Laser photons with energy $h\nu_p$ create a cold distribution of electron-hole carriers. The carriers then heat up by absorbing phonons followed by an up-converted luminescence at $h\nu_f$.

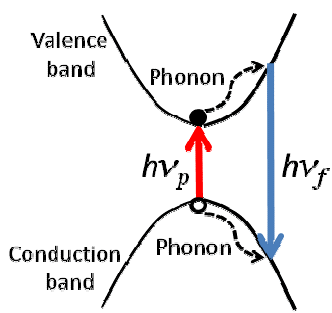

Fig. 3. Cooling cycle in a semiconductor with $h\nu_p$ absorbed energy followed by emission of an up-converted luminescence photon at $h\nu_f$.

Indistinguishable charge carries in Fermi-Dirac distributions allow semiconductors to be cooled to lower temperatures than RE-doped materials. Indeed the highest energy levels of the ground state manifold in the RE-doped systems become less populated as soon as temperature is lowered, due to Boltzmann distribution. The cooling cycle in RE-doped hosts ceases, when the Boltzmann constant times the lattice temperature becomes comparable to the width of the ground state. No such limitations exists in undoped semiconductors. Following to theoretical estimations temperatures as low as 10 K may be achieved in laser cooled semiconductors (Rupper, 2006). Comprehensive theoretical studies of laser cooling of semiconductors has been carried out at Kirtland Air Force Base (Huang et al, 2004 & 2005; Apostolova et al, 2005). The change in the carrier distribution with temperature as well as temperature diffusion of phonons has been calculated. It has been shown that the lattice and the carriers can have differant temperatures varying in space and time. They have suggested multiphoton excitation for pamping wide-band-gap semiconductor (Apostolova et al, 2005). Following to their theoretical estimations the linear excitation scheme seems more favorable for laser cooling. However, nonlinear excitation allows one to use a differant range of photon energies for achieving laser cooling of wide-band-gap semiconductor.

Sheik-Bahae and Epstein (Sheik-Bahae & Epstein, 2004) have developed a theoretical approach to analyse the laser cooling of bulk GaAs based on a microscopic theory for the luminescence and absorption spectra. Let us consider an undoped (intrinsic) semiconductor uniformly irradiated with a laser light at photon energy $h\nu_p$. We assume that only a fraction of the total luminescence can escape the material. It can be described with the extraction efficiency, $\eta_e$, which is the probability that an emitted photon escapes from the semiconductor without being reabsorbed. The remaining fraction (1-$\eta_e$) of the total luminescence is trapped and re-cycled. It contributes to carrier generation. For a given temperature, the rate equation for the electron-hole pair density ($N$) is given by relationship:

$$\frac{dN}{dt} = \frac{\alpha(\nu,N)}{h\nu_p}I - AN - BN^2 - CN^3 + (1-\eta_e)BN^2, \qquad (15)$$

where $\alpha(\nu,N)$ is the interband absorption coefficient. The recombination process consists of nonradiative ($AN$), radiative ($BN^2$), and Auger ($CN^3$) rates. All coefficients are temperature dependent. The last term in Eq. (15) describes increase in the electron-hole pair density caused by re-absorption of the luminiscence that does not escape. We assume that the re-absorption occures within the laser excitation volume. The density-dependence of the interband absorption coefficient $\alpha(\nu,N)$ results from both band-blocking (saturation) and Coulomb screening effects. It can be approximated by a blocking factor (Basu, 1997):

$$\alpha(\nu,N) = \alpha_0(\nu,N)\{f_v - f_c\}, \qquad (16)$$

where $\alpha_0(\nu,N)$ is the unsaturated absorption coefficient, $f_v$ and $f_c$ are Fermi-Dirac distribution functions for the valence and conduction bands, respectively. It is important to emphasise that $\eta_e$ is an averaged quantity over the entire luminescence spectrum.

$$\eta_e = \frac{\int S(\nu)R(\nu)d\nu}{\int R(\nu)d\nu}, \qquad (17)$$

where $S(\nu)$ is the geometry-dependent escape probability of photon ($h\nu$), and $R(\nu)$ is the luminescence spectral density. The radiative recombination can be obtained from:

$$BN^2 = \int R(\nu)d\nu. \qquad (18)$$

The luminiscence spectral dencity is related to the absorption coefficient through Kubo-Martin-Schwinger (KMS) relation (Basu, 1997):

$$R(\nu,N) = \frac{8\pi n^2 \nu^2}{c^2}\alpha(\nu,N)\left\{\frac{f_c(1-f_v)}{f_v - f_c}\right\}, \qquad (19)$$

where $n$ is the refractive index and $c$ is the speed of light.

The net power density ($P_{net}$) deposited in the sample is equal to the difference between the power density absorbed from the pump laser, $P_{abs} = \alpha I + \Delta P$, and that of the escaped luminescence, $P_{lu} = \eta_e B N^2 h\overline{\nu_f}$:

$$P_{net} = \alpha I + \Delta P - \eta_e B N^2 h\overline{\nu_f}. \tag{20}$$

In this equation the first term, $\alpha I$, describes resonant absorption, the second term $\Delta P$ describes undesireble parasitic absorptin such as free-carrier absorption. A mean luminescence energy, $h\overline{\nu_f}$, is defined with the equation:

$$h\overline{\nu_f} = \frac{\int S(\nu)R(\nu)h\nu d\nu}{\int S(\nu)R(\nu)d\nu}. \tag{21}$$

The mean luminescence energy is redshifted from its internal value, $h\nu_f$, which can be calculated with Eq. (21) if the escape probability $S(\nu) = 1$. Using Eq. (15) at steady-state condition, $dN/dt = 0$, one can rewrite (20) in the form:

$$P_{net} = \eta_e B N^2 h(\nu_p - \overline{\nu_f}) + A N h\nu_p + C N^3 h\nu_p + \Delta P. \tag{22}$$

It is obviously that cooling occurs when $P_{net} < 0$. It is important to emphasise that the condition $\nu_p < \overline{\nu_f}$ has to be satisfied and radiative recombination has to dominate (i.e. $\eta_e B/C >> N >> A/\eta_e B$). The cooling efficiency, $\eta_{cool} = P_{cool}/P_{abs}$, is defined as the ratio of cooling power density, $P_{cool} = -P_{net}$, to the absorbed laser power density, $P_{abs} = \alpha I + \Delta P$. Taking into account Eq. (22) and considering (15) at steady-state condition when $dN/dt = 0$, one can obtain the relation for $\eta_{cool}$ in the form:

$$\eta_{cool} = -\frac{\eta_e B N^2 h(\nu_p - \overline{\nu_f}) + A N h\nu_p + C N^3 h\nu_p + \Delta P}{\eta_e B N^2 h\nu_p + A N h\nu_p + C N^3 h\nu_p + \Delta P}. \tag{23}$$

Ignoring parasitic absorption described by $\Delta P$ for the moment one can conclud from Eq. (22) that net cooling can be observed provided that $\eta_e B h(\overline{\nu_f} - \nu_p) \geq 2h\nu\sqrt{AC}$. The cooling efficiency can be presented in more simple form:

$$\eta_{cool} = \eta_{ext} \frac{\overline{\nu_f}}{\nu_p} - 1, \tag{24}$$

where $\eta_{ext}$ is the *external* quantum efficiency desribed by the relation:

$$\eta_{ext} = \frac{\eta_e BN^2}{AN + \eta_e BN^2 CN^3}. \qquad (25)$$

As one can see from (25) there is an optimum carrier density equal to $N_{opt} = (A/C)^{1/2}$ at which external quantum efficiency reaches a maximum:

$$\eta_{ext}^{\max} = 1 - \frac{2\sqrt{AC}}{\eta_e B}. \qquad (26)$$

Taking into account parasitic background absorption $\Delta P = \alpha_b I$, one can obtain the cooling efficiency in more general form:

$$\eta_{cool} = \eta_{abs} \eta_{ext} \frac{\overline{\nu_f}}{\nu_p} - 1, \qquad (27)$$

where the absorption efficiency can be expressed as $\eta_{abs} = \alpha/(\alpha + \alpha_b)$.

Experimental invesigation of laser cooling of semiconductors has been discussed in a number of papers (Gauck et al, 1997; Finkeissen et al, 1999; Eshlaghi et al, 2008). In the first work (Gauck et al, 1997), which was done at the University of Colorado an impressive external quantum effeiciency of 96% in GaAs was achieved, but no net cooling was observed. Later a European consortium announced about local cooling in AlGaAs quantum wells (Finkeissen et al, 1999). This results were later attributed to misinterpretation of spectra caused by Coulomb screening of the excitons (Hasselbeck et al, 2007). Although semiconductors are very promising materials for laser cooling of solids and thier external quantum efficiency increases with decreasing temperature, since the loss terms A and C decrease and the radiative rate (B coefficient) increases inversely with the temperature, there are some problems, which have to be overcome to achieve net cooling of a semiconductor experimentally. (1) *The surface recombination rate has to be reduced*. Well developed epitaxial growth technique such as metal organic chemical vapor deposition (MOCVD), which can provide very low surface recombination rave ($A < 10^4$ sec$^{-1}$) can be considered as a promising solution of the proble. In this case a GaAs active layer is sandwiched between two thin layers of AlGaAs or InGaP. These lattice-matched cladding layers provide surfase passivation and carrier confinement. In the same time (2) *the parasitic background absorption has to be redused* and (3) *the extraction luminescence efficiency has to be enhanced*. The background absorption can be reduced during material preparation with well developed epitaxial methods. The extraction efficiency can be enhanced if total internal reflection, which causes trapping and re-absorption of spontaneous emission, will be prevented. Index matched dome lenses are a solution of the problem. At the present time the purity of the samples is the main obstacle on the way of achievement of the net laser cooling in semiconductors. A novel and very promising method based on the frustrated total internal reflection across a vacuum "nano-gap", where the cooling heterostructuire and luminescence absorber are optically contacted but thermally insulated is under development at the University of New Mexico (Martin et al, 2007).

## 4. Radiation-balanced laser

The processe of excitation and stimulated emission in conventional solid-state lasers results in heat generation in the lasing medium becouse of the Stokes energy shift between the pump photons and the laser output photons commonly called the quantum defect. That is conventional solid-state lasers are always exothermic. Excess heating can cause undesirable changes in laser material such as thermal stress fracture, thermal lensing, and thermal birefringece. It can substantially change the laser gain. These thermal effects limit performance of high-power lasers, where heat loads today usually excess 500 W/cm$^3$. The fiber geometry of the solid-state laser with its very high surface-to-volume ratio providing excellent heat dissipation and thin-disk laser technology have provided tremendous progress in the developmen of high-power fiber devices in recent years. But today the optical intensities in high-power fiber devices have almost reached the damage threshold of the material. Although some lasers with output power as great as several hundreds or thousand watts and with good quality of the beam are developed (Hinea et al, 2000 & Stewen et al, 2000) thermal effects limit conventional bulk and fiber solid state lasers to reach higher power levels.

In 1999, Bowman suggested a new solid-state bulk laser design without internal heat generation called a *radiation balanced* or *athermal* laser (Bowman, 1999). He suggested the use of radiation cooling by anti-Stokes fluorescence within the laser medium to balance the heat generated by the Stokes shifted stimulated emission in a solid-state bulk laser. According to Bowman's concept processes of lasing and anti-Stokes cooling occur in the same system of ions. In 2001, Andrianov and Samartsev (Andrianov & Samartsev, 2001) proposed a scheme, where lasing occurs in one system of ions, while anti-Stokes cooling takes place in another system of ions. In 2007, a nover solution to reduce heat generation in powerful optically pumped Raman lasers was propoused (Vermeulen et al, 2007). A nonlinear phenomenon called coherent anti-Stokes Raman scatterimg (CARS) was used to minimize the heat generated by the quantum defect phenomenon. In 2009, schemes of athermal amplifiers with „integrated" coolers were also proposed (Nemova, 2009; Nemova & Kashyap, 2009).

### 4.1 Radiation-balanced lasing

Consider basic concepts of radiation balanced (athermal) laser, when lasing and anti-Stokes cooling occurs in the same system of ions doped in the crystal or glass host. Fig. 4 illustrates the energy-level diagram of a quasi-two-level laser system, where quantum energy defect is only of the order kT. Solid-state laser of this type can be often referred to as quasi-three-level laser. The upper and lower elecronic levels (manifolds) are splited into many closely spaced sublevels. The population of each sublevel within a manifold is described by Boltzmann occupation factors. We assume that transitions between these sublevels are purely nonradiative transitions, provided by phonon absorption and emission. The energy gap between sublevels is much less that kT thus intra-band thermalization occurs on a picosecond time scale. Assume that radiative lifetime of the upper manifold is on the order of milliseconds and transitions between the upper and lower manifolds (inter-band relaxation) are purely radiative, since a bandgap between the upper and lower manifolds is large compared to the energies of the phonons. We also assume the absence of excited-state absorption, energy transfer, and absorption by nonradiative background transitions. The laser is pumped at the frequency $v_p$. $v_l$ is the frequency of the laser field, and $v_f$ is mean fluorescence frequency. The total density of the ions in the host, $N_T$, is equal to the sum of the densities of ions in the first, $N_1$, and second, $N_2$, energy levels:

$$N_T = N_1 + N_2. \tag{28}$$

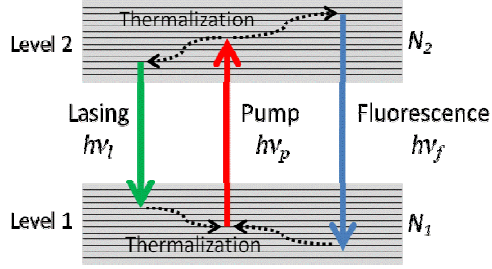

Fig. 4. Energy-level scheme for a radiation-balanced laser.

The rate equation of the upper level is

$$\frac{dN_2}{dt} = W_p - W_l - \frac{N_2}{\tau}, \tag{29}$$

where $\tau$ is the fluorescence lifetime, $W_p$ is a pump rate, and $W_s$ is a stimulated emission rate described by the formulas:

$$W_p = \frac{I_p}{h v_p}\left[N_T \sigma_p^a - N_2\left(\sigma_p^a + \sigma_p^e\right)\right],$$
$$W_l = \frac{I_l}{h v_l}\left[N_2\left(\sigma_l^a + \sigma_l^e\right) - N_T \sigma_l^a\right], \tag{30}$$

where $I_{p,l}$ are intensities of the pump ($p$) and the laser ($l$) beams. $\sigma_{p,l}^{a,e}$ are the cross sections of the absorption ($a$) and stimulated emission ($e$) of the pump ($p$) and the laser ($l$) wavelengths. In the steady state, $dN_2/dt = 0$ and equation (29) can be written as

$$W_p = W_l + \frac{N_2}{\tau}. \tag{31}$$

It is important to emphasise that for radiation-balanced amplification, the absorbed power density has to be equal to the radiated power density at any point in the laser mediuam:

$$h\nu_p W_p = h\nu_l W_l + h\nu_f \frac{N_2}{\tau}. \tag{32}$$

The equation (32) is inherent for athermal laser only, and is *not valid* in a traditional exothermic laser in which the Stokes energy shift between the pump photons ($h\nu_p$) and the laser photons ($h\nu_l$) appears as heat in the amplifier medium. The relation for laser gain can be described by the well known equation

$$\frac{dP_l}{dz} = \left[\left(\sigma_l^a + \sigma_l^e\right)N_2 - \sigma_s^a N_T\right]P_l. \tag{33}$$

Substituting Eqs. (31) and (32) in Eq. (33) one can obtain the equation, which describes the laser signal at any point, $z$, along the length of the laser medium.

$$\frac{P_l(0)}{P_l(z)}\exp\left(\frac{P_l(z)-P_l(0)}{P_l^{Sat}}\right) = \exp(\sigma_l^a N_T z), \tag{34}$$

where

$$P_l^{Sat} = A_{eff}\frac{h\nu_l}{\tau\left(\sigma_l^a+\sigma_l^e\right)}\left(\frac{\nu_f-\nu_p}{\nu_p-\nu_l}\right) \tag{35}$$

is the saturation power of the laser signal and $A_{eff}$ is the effective area of the mode, which supports laser signal. To support grow of the laser signal for one-way propagation described by Eq. (34) and keep radiation balance at each point in the laser medium the pump power has to be properly distributed along the length of the laser medium. This distribution can be obtained with Eqs. (30)-(32):

$$P_p(z) = \frac{\sigma_l^a\left(\sigma_p^a+\sigma_p^e\right)P_l(z)P_p^{Sat}}{\left(\sigma_p^a\sigma_l^e - \sigma_l^a\sigma_p^e\right)P_l(z) - \sigma_p^a\left(\sigma_l^a+\sigma_l^e\right)P_l^{Sat}}, \tag{36}$$

and

$$P_p^{Sat} = A_{eff}\frac{h\nu_p}{\tau\left(\sigma_p^a+\sigma_p^e\right)}\left(\frac{\nu_f-\nu_l}{\nu_p-\nu_l}\right). \tag{37}$$

It is easily seen from Eq. (36) that radiation-balanced amplification requires careful control of the pump power distribution along the laser medium. As can be seen from Eq. (36), since the value of the pump power has to be $P_p > 0$, in the case of athermal laser for each combination of the host material, ions, pump and signal wavelengths there is a minimum injected laser power:

$$P_l^{\min} = A_{eff} \frac{h\nu_l \sigma_p^a}{\tau(\sigma_p^a \sigma_l^e - \sigma_l^a \sigma_p^e)} \left(\frac{\nu_f - \nu_p}{\nu_p - \nu_l}\right). \tag{38}$$

As one can see from Eqs. (31) and (32) for radiation-balanced laser operation the mean fluorescence frequency, the pump and laser frequencies have to satisfy to the relation: $\nu_f > \nu_p > \nu_l$.

Contrary to traditional exothermic laser the power of the amplified signal changes almost linearly with the length of the fiber in the case of athermal amplification. Analysis of sensitivity and stability of a radiation-balanced laser to perturbations in the field parameters and temperature was done in the paper (Bowman, 2002). It has been shown that fluctuations in the gain set limits on the variability of the pump wavelength. A pump stability of ±1 nm is suggested for Yb:KGW laser. As active wavelength stabilization scheme is proposed to minimize sensitivity of the athermal laser to ambient temperature fluctuations. Thermodynamics of radiation-balanced laser has been comprehensively analysed by Mungan (Mungan, 2003). Carnot efficiency has been derived for all-optical amplification from consideration of the radiative transport of energy and entropy. The highest Carnot efficiencies result only when the system is pumped into saturation. In 2002, Bowman and his coworkers experimentally demonstrated the first athermal laser (Bowman, 2002). In this laser only 0.42% of the pump power absorbed by the medium was converted into heat.

**4.2 Self-cooling laser**
In this paragraph we will consider basic concepts of a self-cooled laser, where lasing occurs in one system of ions, while anti-Stokes cooling takes place in another system of ions co-doped in the laser host. An idea of a self-cooled laser was proposed in 2001 (Andrianov & Samartsev, 2001). It was suggested to co-dope $Nd^{3+}$:$KY_3F_{10}$ laser with $Yb^{3+}$ ions providing anti-Stkes cooling. Enerdy levels of the $Nd^{3+}$ and $Yb^{3+}$ in $KY_3F_{10}$ crystal is illustrared in Fig. 5. In the case of CW pump power exceeding the value of the laser threshold, a coherent electric field with high intensity accumulated in a laser cavity and corresponding to the transition between levels 1 and 2 of $Nd^{3+}$ ion compesates all the losses related to both radiation coupling out of the cavity and the outgoing flow of field energy to $Yb^{3+}$ ions responsoble for cooling. The lasing wavelength falls within the long-wavelength wing of the absorption line of ytterbium and small fraction of the lasing beam serves as a pump for $Yb^{3+}$ ions. Losses through $Yb^{3+}$ ions are mainly determined by the rate of its spontaneous luminescence. The heating of a laser is predominantly due to nonradiative transitions of $Nd^{3+}$ ions from levels $^3F_{5/2}$ -> $^3F_{3/2}$ creating phonons with energies $\hbar\Omega_{32}$. The splitting $\hbar\Omega_{32}$ is two to three times less than the splitting $\hbar\Omega_{o'1'}$ of the ground state of $Yb^{3+}$ ions resulting in cooling. It has been shown (Petrushkin et al, 2002 & 2003) that when the concentrations of $Nd^{3+}$ and $Yb^{3+}$ ions are of the same order of magnitude then the cooling power is approximately four times greater than the pumping-induced heating power. The lasing threshold condition for the population difference in the presence of $Yb^{3+}$ ions requires larger values of population inversion than the threshold in the absence of co-doped $Yb^{3+}$ ions.

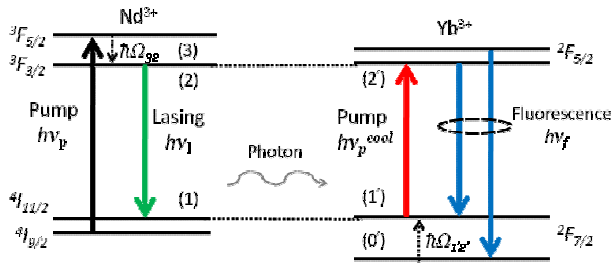

Fig. 5. Energy-level scheme for a self-cooling laser.

Self-cooling of a four-level solid-state laser has been analysed too (Andrianov et al, 2004). It is shown that such lasers cannot be build with traditional active media. They can be released employing materials allowing a large frequency detuning upon cooling or with schemes in which no heat is released upon pumping.

## 4.3 CARS-based heat mitigation in Raman lasers

High power Raman lasers, the lasing mechanism of which is based on stimulated Stokes Raman scattering (SSRS) in a Raman medium and generate wavelengths beyond the reach of other lasers. Like all traditional lasers they suffer from the heat dissipation inside the active medium caused by the quantum defect between the pump and lasing (Stokes) photons, which deteriorates the performance on the laser. As shown in previous paragraphs the main requirement for solids to allow efficient anti-Stokes fluorescence cooling is the requirement that the average frequency of the fluorescence spectrum has to be higher than the low-frequency limit of their absorption spectrum. Raman media do not feature a fluorescence band and neither anti-Stokes fluorescence cooling nor radiation-balanced laser can be realised in these media. A new approach permitting to mitigate the heat dissipation in Raman lasers has been proposed (Vermeulen et al, 2007). This intrinsic heat-mitigation technique relies on coherent anti-Stokes Raman scattering (CARS) instead of anti-Stokes fluorescence. It relies on three different Raman processes that can occur inside the medium of an optically pumped Raman laser illustrated in Fig. 6. These are stimulated Stokes Raman scattering (SSRS), stimulated anti-Stokes raman scattering (SARS), and coherent anti-Stokes Raman scattering (CARS). The maximum Raman gain point is considered here, that is all scattering proocess are exactly at Raman resonance. The SSRS is actual lasing mechanism: an incoming pump photon scatters onto an internal material osillatin. The SSRS result is creation of a lower energy Stokes photon and generation of a phonon of heat, i. e., quantum-defect heatingt. The SARS process generates quantum-defect heating like SSRS process: as anti-Stokes photon is converted into a pump photon and a phonon Fig. 6 (b). The CARS at Raman resonance illustrated in Fig. 6(c) is a four-wave mixing process that converts a pump photon and a Stokes photon into a pump photon and an anti-Stokes photon, while annihilating two phonons (Bobbs, 1990). This process reduces the quantum-defect heating in the Raman medium. In case of a phase mismatch, the induces phase variation along the propagation way of the field will cause the CARS process to alternate with reverse mechanism. This point of view developed by Bobbs (Bobbs, 1990) is different from traditional one, where the CARS is mechanism that coverts two pump photons into a Stokes ans anti-Stokes photons without exchanging energy with the Raman medium.

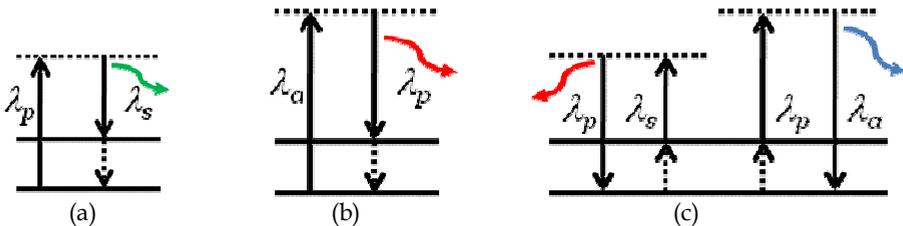

Fig. 6. Prodesses, which take place in a Raman laser: (a) stimulated Stokes Raman scattering (SSRS), (b) stimulated anti-Stokes raman scattering (SARS), (c) coherent anti-Stokes Raman scattering (CARS).

The efficiency of CARS-based heat mitigation in the Raman laser can be esstimated using Stokes – anti-Stokes iterative resonator method (Vermeulen et al, 2006), which enable one to describe the steady-state regime and the transient phenomena of a CW pumped Raman laser emitting both Stokes and anti-Stokes scattered photons. Applying for a certain half round-trip (α) the method to the amplitudes of the intra-cavity pump fields $E^+_{p,(\alpha)}(z)$ and $E^-_{p,(\alpha)}(z)$, Stokes fields $E^+_{s,(\alpha)}(z)$ and $E^-_{s,(\alpha)}(z)$, and anti-Stokes fields $E^+_{a,(\alpha)}(z)$ and $E^-_{a,(\alpha)}(z)$, propagating in the forward and the backward direction, respectively, yields:

$$\frac{\partial E^{\pm}_{p,(\alpha)}}{\partial z} = \mp \left[ \frac{\lambda_s}{\lambda_p} \left( G^{\pm}_s \left| E^+_{s,(\alpha)} \right|^2 + G^{\mp}_s \left| E^-_{s,(\alpha)} \right|^2 \right) \right] E^{\pm}_{p,(\alpha)} \\ \pm \left[ \left( G^{\pm}_p \left| E^+_{a,(\alpha)} \right|^2 + G^{\mp}_p \left| E^-_{a,(\alpha)} \right|^2 \right) \right] E^{\pm}_{p,(\alpha)} - \gamma^{\pm}_p E^{\pm}_{p,(\alpha)},$$

(39)

$$\frac{\partial E^{\pm}_{s,(\alpha)}}{\partial z} = \pm \left[ \left( G^{\pm}_s \left| E^+_{p,(\alpha)} \right|^2 + G^{\mp}_s \left| E^-_{s,(\alpha)} \right|^2 \right) \right] E^{\pm}_{s,(\alpha)} \\ \pm \frac{\lambda_a}{\lambda_s} C_{sa} \left( E^{\pm}_{p,(\alpha)} \right)^2 \left( E^{\pm}_{a,(\alpha)} \right)^* e^{\pm i\Delta k(z+\Lambda)} - \gamma^{\pm}_s E^{\pm}_{s,(\alpha)},$$

(40)

$$\frac{\partial E^{\pm}_{a,(\alpha)}}{\partial z} = \mp \left[ \frac{\lambda_p}{\lambda_a} \left( G^{\pm}_p \left| E^+_{p,(\alpha)} \right|^2 + G^{\mp}_p \left| E^-_{p,(\alpha)} \right|^2 \right) \right] E^{\pm}_{a,(\alpha)} \\ \mp C_{sa} \left( E^{\pm}_{p,(\alpha)} \right)^2 \left( E^{\pm}_{s,(\alpha)} \right)^* e^{\pm i\Delta k(z+\Lambda)} - \gamma^{\pm}_a E^{\pm}_{a,(\alpha)},$$

(41)

where $G_s^{\pm}$ are the SSRS Stokes gain coefficients for forward (+) and backward (-) SSRS, $G_p^{\pm}$ are SARS pump gain coefficients for forward (+) and backward (-) SARS, $C_{sa}$ is CARS coupling coefficient for a pump linewidth $\Delta\lambda_p$ within the limits set by dispersion. These limits determine the pump linewidth for which the pump field still coherently propagates with the Stokes and anti-Stokes fields but onlyin the co-propagating dcattering direction. $\lambda_s$

and $\lambda_a$ are the Stokes and anti-Stokes wavelengths, and $\lambda_p$ is the pump wavelength. The output field amplitudes at the half round-trip ($\alpha$) throught the front and back cavity mirrors

$$G_s^+ = \frac{g}{4}\sqrt{\frac{\varepsilon_0}{\mu_0}}, \qquad G_p^+ = \frac{\lambda_s}{\lambda_p}G_s^+, \qquad C_{sa} = \frac{\lambda_s}{\lambda_a}G_s^+, \qquad G_{s,p}^- = \frac{\Delta\nu_R G_{s,p}^+}{\Delta\nu_R + \Delta\nu_p}. \qquad (42)$$

here, $g$ is the Raman gain coefficient and $\Delta\nu_R$ is the spontaneous Raman linewidth. $\Delta k = 2k_p - k_s - k_a$ is phase mismatch of the CARS process, $z$ varies between $z = 0$ and $z = L$, whereas $\Lambda$ equals the total distance that the field have traversed inside the cavity over all previous half round trips. The set of equations (39) – (41) have to be completed by the boundary conditions at the front and back cavity mirrors, which couples successive half round trips with the output fiel amplitudes $E^{f,b}_{p,s,a,(\alpha)}$.

$$\begin{aligned}
E^f_{p,(\alpha)} &= \sqrt{R_{p,f}}E^{in}_{p,(\alpha)} - \sqrt{T_{p,f}}E^-_{p,(\alpha)}(0)e^{i\delta_p}, \\
E^b_{p,(\alpha)} &= -\sqrt{T_{p,b}}E^+_{p,(\alpha)}(L), \\
E^f_{s,(\alpha)} &= -\sqrt{T_{s,f}}E^-_{s,(\alpha)}(0), \\
E^b_{s,(\alpha)} &= -\sqrt{T_{s,b}}E^+_{s,(\alpha)}(L), \\
E^f_{a,(\alpha)} &= -\sqrt{T_{a,f}}E^-_{a,(\alpha)}(0), \\
E^b_{a,(\alpha)} &= -\sqrt{T_{a,b}}E^+_{a,(\alpha)}(L),
\end{aligned} \qquad (43)$$

here $\delta_p = 4\pi(L/\lambda_p)n_p - 2\pi m$, where $n_p$ is the refractive index at $\lambda_p$ wavelength and $m$ is an integer. The CARS-based heat-mitigation efficiency $\eta_{HM}$ the ration of the number of outcoupled anti-Stokes photons to the number of outcoupling Stokes photons:

$$\eta_{HM} = \frac{\left|E^f_{a,(\alpha)}\right|^2 + \left|E^b_{a,(\alpha)}\right|^2 + \left|E^{loss}_{a,(\alpha)}\right|^2}{\left|E^f_{s,(\alpha)}\right|^2 + \left|E^b_{s,(\alpha)}\right|^2 + \left|E^{loss}_{s,(\alpha)}\right|^2}\frac{\lambda_a}{\lambda_s} \times 100\%. \qquad (44)$$

The total power losses integrated over the cavity length at half round-trip ($\alpha$) are represented by the electric field amplitudes $E^{loss}_{s,a,(\alpha)}$. In order to increase CARS-based heat-mitigation efficiency more anti-Stokes photons have to be generated. Effective heat mitigation can be obtained by realizing the phase matching for CARS process, $\Delta k = 0$, and by reducing the backward Raman scattering. For pump linewidth within the dispersion limits the gain of backward Raman scattering will decrease with increasing pump linewidth whereas the gain of forward Raman scattering will remain unchanged.

The CARS-based heat mitigation technique has be demonstrated for two differant types of Raman lasers (Vermeulen et al, 2007): a $H_2$-based Raman laser (Benabid et al, 2005), and Si-based Raman laser (Dimitropoulos et al, 2004). The $H_2$-based Raman laser consists of a hollow-core photonic crystal fiber (HC PCF) filed with $H_2$ and spliced at both sides to standard fibers with Bragg gratings. We assume that phase matching, $\Delta k = 0$, can be obtained in the HC PCF (Konorov et al, 2005). The Si-based Raman laser is silicon-on-insulator waveguide Raman laser, where the phase matching condidtion $\Delta k = 0$ can be obtained. It has been obtained that the CARS-based heat mitigation mechanism for Raman lasers can reduce the average quantum-defect heating per out-coupled Stokes photon by at least 30% and 35% for a $H_2$- and a Si-based Raman lasers, respectively. One potential drawback of the CARS-based heat mitigation technique is linear growth of the radiated energy with the length of the mediumin in phase-matched CARS. In traditional Raman lasers it increases exponentially with the of the mediumin.

## 5. Conclusion

At the present time cooling of RE-doped materials makes rapid strides and approaches cryogenis operation. In the nearest years, the advantageous of optical refrigeration should be beneficial for satellite instrumentations and small sensors, where compactness and the lack of vibrations are very important. The cooling element is not electrically powered. It will not interfere with the electronics, which will be cooled. It can be very useful in the production electronics incorporating superconducting components. Optical refrigeration can be used for development of magnetometers for geophysical and biomedical sensing. The purity of the materials is crucial parameter for its practical realization of optical refrigerator. In the recent years much progress has been made in achieving high external quantum efficiency in semiconductors. We hope with the advanced heterostructure growth cooling in semiconductors will soon be attainable.

Future research on a way of development of optical refrigerator has to be concentrated on new host materials with low phonon energies. Other dopants have to be investigated. The optimization of the geometry of the refrigerating material and effective removal or recycling of the fluorescence has to be carefully explored. In ideal case, the fluorescence has to be recycled into the laser input beam by some down-conversion process. Improvement of the beam quality of the pump laser is very important for high efficient refrigeration. Using photonic crystals as hosts for RE ions opens new possibility of enhancing the absorption and emission of light because the optical properties of the system can be tailored for specific applications. For example, the enhancement of the absorption rate can be achieved by adjusting geometrical parameters due to the non-trivial character of the density of photonic states in photonic crystals.

Radiation-balanced laser, which can be realized with different schemes, is a promising solution for development high power lasers and amplifiers. They permit to get very high output powers with high quality laser beam.

25